\documentclass[12pt]{article}

\usepackage[utf8]{inputenc}
\usepackage{roboto}

\usepackage{setspace}
\usepackage{lscape}
\usepackage{amsmath}
\usepackage{amsthm}
\usepackage[bottom=1in,top=1in,left=1in,right=1in]{geometry}
\usepackage{indentfirst}
\usepackage[colorinlistoftodos]{todonotes}
\usepackage{url}

\usepackage{titlesec}
\titlespacing*{\chapter}{0pt}{-15pt}{40pt}
\titlespacing{\section}{0pt}{0cm}{0cm}
\titlespacing{\subsection}{0pt}{0cm}{0cm}
\usepackage{appendix}

\usepackage{multirow}

\usepackage{enumerate}
\usepackage[shortlabels]{enumitem}
\setlist[itemize]{topsep=0pt}
\setlist[enumerate]{topsep=0pt}

\usepackage{colortbl}
\usepackage{xcolor}
\definecolor{blue}{RGB}{0,0,153}
\definecolor{red}{RGB}{204,0,0}
\definecolor{yellow}{RGB}{255,255,153}
\definecolor{green}{RGB}{0,102,0}
\definecolor{orange}{RGB}{255,128,0}
\definecolor{purple}{RGB}{102,0,204}
\definecolor{orange_graph}{RGB}{245, 121, 58}
\definecolor{blue_graph}{RGB}{15, 32, 128}

\newcolumntype{a}{>{\columncolor{blue!15}}c}

\usepackage{amssymb,amsfonts,amsthm}
\usepackage{mathtools}
\usepackage{bbm}

\usepackage[multiple]{footmisc}

\usepackage{graphicx,graphics,url,caption}
\usepackage{subcaption} 
\usepackage{float}
	
\usepackage{tikz}
\usetikzlibrary{decorations.pathreplacing,angles,quotes}
\usetikzlibrary{patterns}
\usepackage{array,booktabs,arydshln}

\usepackage{adjustbox}
\usepackage[flushleft]{threeparttable}

\usepackage{natbib}
\setcitestyle{citesep={,}}

\usepackage{hyperref}

\usepackage{siunitx}
\newcolumntype{d}{S[input-symbols = ()]}

\usepackage{listings}
\usepackage{color} 
\definecolor{mygreen}{RGB}{28,172,0} 
\definecolor{mylilas}{RGB}{170,55,241}

\usepackage{booktabs}
\usepackage{siunitx}
\newcolumntype{d}{S[input-symbols = ()]}


\begin{document}

\title{\vspace{-1cm}School closures and educational path: how the Covid-19 pandemic affected transitions to college\footnote{ Financial support from the Sao Paulo Research Foundation (FAPESP - grant \# 2015/21640-3, 2017/50134-4 and 2019/25033-5) and CNPq are gratefully acknowledged. We thank COMVEST, UNICAMP's admission office, for providing the data and assistance during the project. This research has benefited from discussions with Rodolfo Jardim de Azevedo and Rafael Maia. We also thank Luis Felipe Lapa Barcelos Coutinho, Larissa Gomes de Stefano Escaliante, Gabriel Braga Proença, Gabriel Gomes de Siqueira, and Fernando Santos de Souza for their excellent research assistance. All remaining errors are ours.}}
\author{Fernanda Estevan\footnote{Sao Paulo School of Economics - FGV,  fernanda.estevan@fgv.br} \and Lucas Finamor\footnote{Yale University --- Department of Economics, lucas.finamor@yale.edu.}}


\maketitle

\pagenumbering{arabic}  

\begin{abstract}
\noindent We investigate the impact of the Covid-19 pandemic on the transition between high school and college in Brazil. Using microdata from the universe of students that applied to a selective university, we document how the Covid-19 shock increased enrollment for students in the top 10\% high-quality public and private high schools. This increase comes at the expense of graduates from relatively lower-quality schools. Furthermore, this effect is entirely driven by applicants who were at high school during the Covid pandemic. The effect is large and completely offsets the gains in student background diversity achieved by a bold quota policy implemented years before Covid. These results suggest that, not only students from underprivileged backgrounds endured larger negative effects on learning during the pandemic, but they also experienced a stall in their educational paths.

\end{abstract}


\doublespacing

\clearpage
\section{Introduction}
The Covid-19 pandemic has severely affected education systems across the globe. School closures with the sudden transition to online education were disruptive, and the negative effects on learning have been documented from primary to tertiary education.\footnote{The evidence for primary and secondary education has been documented by \cite{asanov2021,engzell2021,grewenig2021,lichand2022,maldonado2022}. Several papers also show the effects on college students, among them \cite{aucejo2020,bacher2021,bulman2022,kofoed2021,rodriguezplanas2022,failache2022}.}
Several studies documented the reduction in educational inputs, such as the time spent studying \citep{asanov2021,grewenig2021}, changes in students expectations and beliefs \citep{aucejo2020} and worse outcomes in performance \citep{kofoed2021,engzell2021,lichand2022,maldonado2022}. As expected, these effects were larger for students from disadvantaged backgrounds, amplifying socioeconomic achievement disparities. 

However, we know less about how the pandemic and its negative educational outcomes affected critical educational transitions, such as the passage from high school to university. This question is particularly relevant for developing countries where the proportion of low SES students reaching college is low, despite evidence of substantial college premiums, particularly in selective universities \citep[e.g.,][]{zimmerman14,sekhri20,jia21}.

In this paper, we investigate the impact of the Covid pandemic on the transition from high school to university, exploring the admissions to a selective university in Brazil, Unicamp. Unlike most Brazilian public universities, Unicamp manages its admission system, allowing us to observe all applicants to its admission processes. As a result, we have access to a rich array of applicants' socioeconomic characteristics and exam performance data. Crucially, our dataset includes applicants' high school names, allowing us to evaluate how the Covid pandemic affected applicants' and enrollees' educational profiles, considering the critical transition from high school to college.

We construct a high-school quality measure using publicly available data from ENEM, an end-of-high-school exam widely used for college admission in Brazil, for every high school in the country. Then, merging the high-school quality data with our admissions data, we can identify the percentiles of the high-school quality distribution from which Unicamp applicants and enrollees are drawn before and after the onset of the Covid pandemic.

Since Unicamp's admission system has experienced significant changes over the years, we focus our analysis on three crucial periods surrounding the Covid pandemic. The first period ranges between 2016 and 2018 when the Unicamp admission system only relied on its regular exam. During that period, Unicamp used an affirmative action policy that granted bonus points to public high school graduates and underrepresented racial minorities in the admission exam. Then, starting in 2019, Unicamp explicitly implemented quotas for public high school graduates and racial groups by reserving a proportion of its seats and opening up new admission processes. For our purposes, the quota period provides a useful benchmark to compare the magnitude of the Covid pandemic impact relative to a policy designed to facilitate underprivileged students' access to university. Finally, our main focus is on the 2021 and 2022 admission processes that occurred after the beginning of the Covid pandemic.

We start by documenting the existence of a small set of high-quality schools that account for a large proportion of Unicamp applications and enrollments, even before the pandemic. Schools in the top 10\% have only 7.2\% of students in the country and 62\% of enrolled students at Unicamp. Those in the bottom 70\% of the school quality distribution account for 80.9\% of high-school students in the country and only 16.5\% of students enrolled at Unicamp. One in every 180 students in a top 10\% school is enrolled at Unicamp, while this proportion is one in 313,589 students in a bottom 10\% school. The best 100 private and 100 public schools teach 0.8\% students and correspond to 15\% of enrolled students. 

We then document how introducing a bold affirmative action policy in 2019 helped reduce this concentration from high-quality schools. The quota policy reduced the probability that an enrolled student comes from the top 10\% by 5.1 percentage points (pp), increasing the chances of students in the bottom 40\% by 1.9pp (a 70\% increase) and for students from schools in the 70th-90th percentiles (by 2.6pp or 12\%). The quota also reduced the probability of a student coming from the best 100 private and 100 public schools from 15\% to 12.3\%, an almost 20\% reduction.

Finally, we show how the Covid pandemic reverted the deconcentration effects achieved by the quota policy. In the two admission processes after the pandemic, the probability of an admitted student being drawn from a top 10\% school increased by 5.7pp, erasing all the gains achieved by the quota policy. This increase came at the expense of students in the entire distribution, particularly in median schools, between the 40th-70th percentile. 

To test whether these effects were related to school closures and the impact of the pandemic on learning, we estimate the Covid effect separately for students enrolled in high school during the pandemic and students that had already graduated in 2019. We show how the effect is entirely driven by students that were in high school during the pandemic. This result suggests that school shutdowns and online classes had an important role in the transition to college. 

We perform several additional exercises to assess the robustness of these results. We first show how these results are qualitatively the same if we instead use different quality measures that take into account the public and private high school disparities. We also show that we obtain the same qualitative results if we use proxies for socioeconomic status at the individual level, such as race, mother education, and family income, instead of our high-school quality measure. Lastly, we show that secular trends in the outcomes do not drive the effects of the quota and Covid. 

Thus, our results point to persistent negative effects of the Covid pandemic. Indeed, not only students from underprivileged backgrounds endured larger negative effects on learning during the pandemic, but they also experienced a stall in their educational paths, affecting their access to selective universities. Moreover, given the large monetary returns associated with selective universities \citep[e.g.,][]{sekhri20}, these negative shocks possibly translate into decreased social mobility.

More broadly, our paper complements the increasing literature reporting the widespread impacts of the Covid pandemic on the educational system and its disproportionate burden on low SES college students. Most of the literature focuses on college student's performance, as measured by delayed graduation and reductions in retention rates \citep{aucejo2020,rodriguezplanas2022,failache2022}. 

Instead, we document the impact of the Covid pandemic on the transition from high school to college. A close paper is \cite{bulman2022} that shows the enrollment reductions in community colleges in California during the pandemic. As in our paper, they show that enrollment also fell more for disadvantaged groups such as African Americans and Latinos. However, our context is quite different as we focus on a selective university in a developing country. Not only are university access inequalities more pronounced, but the pandemic's impact on the school system was probably more pervasive. Accordingly, our results show that the pandemic's negative shocks may further hinder the already turbulent path of low SES students into selective universities in developing countries. 

Our paper also relates to the literature investigating the reasons behind the college application gaps between low and high-income individuals \citep{hoxby13,dynarski21}. As in \cite{hoxby13}, our results point to the crucial role of selective high schools in improving disadvantaged students' chances of getting into college. Furthermore, we show that high-quality schools can safeguard against negative shocks, such as the Covid pandemic, in transitioning from high school to university.

Finally, we also contribute to the literature that exploits the impact of high-quality schools on later outcomes. For example, \cite{berkowitz11} highlights the important role that high-quality schools play in admission to selective universities in the US. In contrast, \cite{dobbie14} finds no impact of having higher-achieving peers on college enrollment, using data from three selective high schools in New York City.

We organize our paper as follows. Section \ref{section:institutional} describes the institutional features of the Brazilian educational system and Unicamp's admission process. Section \ref{section:data} describes our dataset and explains our high-school quality measure. Section \ref{section:results} presents our main results and some robustness checks. Finally, Section \ref{section:conclusion} concludes.

\section{Institutional Setting} \label{section:institutional}

\subsection{Brazilian Educational System}

In Brazil, students enroll in high school for three years in public or private schools. Public high schools accounted for 87.5\% of the enrolled students in 2019 \citep{inep21}. Public schools are tuition-free and funded by the government. Private schools are independently managed and are free to set tuition. According to several indicators, including national standardized exams, public high schools lag behind private schools in proficiency. 

The transition from high school to college is through an admission process that can rely on institute-specific exams.\footnote{ In recent years, there has been a consolidation movement on using a centralized admission process for federal (public) universities. For more details, see \cite{machado2021} and \cite{otero2021}.} As in many countries, students choose majors during the admission process. Higher education institutions can also be public or privately managed. However, in sharp contrast to primary and secondary education, public universities outperform private universities. Public universities are large research institutes administered by the federal or state governments. College length varies from three to six years, depending on specific majors and institutions. 

Since 1998, students may take an end-of-high-school exam called ENEM. Up to 2008, students could use the ENEM score as part of the admission procedure in some universities and colleges. The exam was also required for access to public student loans and scholarships. In 2009, the ENEM exam was reformulated and expanded. Since then, students have taken the exam in four different areas (humanities, science, math, and language) in addition to a written essay. The reformulated exam is constructed using the item response theory (IRT), making the scores comparable across years.
Several universities started using the ENEM exam as their sole admission process with the new version. The take-up for the exam has been high since it became high-stakes for high school graduates. In our baseline period, from 2016-2018, the average take-up rate across all high schools is 74.2\%.\footnote{To compute the take-up rate, we use the number of students enrolled in the ENEM exam in the school years 2015, 2016, and 2017 (corresponding to the admission years 2016-2018) who declared they were graduating from high school and the number of students in the senior year of high school in the School Census.} 

\subsection{Unicamp}

The \textit{Universidade Estadual de Campinas}, Unicamp, is a selective research-intensive public (state) university located in Campinas, Sao Paulo state.\footnote{ For instance, the Times Higher Education ranks Unicamp third in the 2022 Latin America University Rankings \citep{the22}} Unicamp offers undergraduate (and graduate) degrees in several fields of study and does not charge tuition fees like most Brazilian public universities. As is typically the case in Brazilian universities, individuals apply for a specific undergraduate degree.\footnote{ Applicants can also choose a second major. However, since Unicamp uses a version of the Boston mechanism to allocate students to majors, nearly all applicants obtain admission to their first choice \citep{estevan19b}.} While most Unicamp applicants (around 85\% in 2016-2021) come from the Sao Paulo state, Unicamp also draws part of its student population from other states. Consequently, admission to Unicamp, particularly to some majors, is highly competitive. The acceptance rate is around 7.7\% for our baseline period, 2016-2018, but it can be as low as 1.1\% in Medicine, its most competitive degree.

Between 1987 and 2018, nearly all prospective students wrote the Unicamp admission exam, virtually the only path to admission.\footnote{ Starting in 2011, Unicamp also offered 120 slots to the best-ranked students of Campinas' high schools through ProFIS (\textit{Programa de Formação Interdisciplinar Superior}). We do not have data on ProFIS applicants.} The Unicamp's admission exam consists of Phases 1 and 2. In each phase, applicants must answer questions on typical high school subjects.\footnote{ The high school subjects are Portuguese, history, geography, mathematics, physics, chemistry, biology, and English. Until 2010, both phases consisted of open-ended questions. From 2011 on, Phase 1 questions are multiple-choice.} Phase 1 exam is qualifying, and the final admission score combines Phases 1, 2, and ENEM (end-of-high school) scores. In a nutshell, Unicamp ranks applicants by major in decreasing order of final scores and sends admission offers to the best-ranked applicants.

In 2005, Unicamp implemented an affirmative action policy, \textit{Programa de Ação Afirmativa e Inclusão Social} (PAAIS), to address the historical underrepresentation of public high school students and ethnic minorities at Brazilian public universities. As a result, public high school students were eligible to apply through PAAIS and received bonus points in the admission exam.\footnote{ Starting in 2005, Unicamp granted 30 points in Phase 2 to public high school graduates and ten additional points if they self-declared Black or Native. Unicamp doubled both bonuses in 2014. See \cite{estevan19a} for more details on Unicamp's admission process and affirmative action policy.} Between 2016 and 2018, Unicamp granted bonus points for public high school students in Phase 1 (60 points) and 2 (90 points). If they also self-declared Black or Native, the bonus points increased to 80 and 120 points in Phases 1 and 2, respectively.\footnote{The bonuses are added to the normalized score, which has a mean of 500 points and a standard deviation of 100.} Among the students enrolled at Unicamp in 2016-2018, 48,9\% received PAAIS bonuses.


In 2019, Unicamp's affirmative action program underwent significant changes with the introduction of quotas (i.e., reserved seats) for Blacks and public high school graduates. Eligible applicants could apply through the racial quotas in two different ways. First, they could opt for racial quotas upon registering for Unicamp's admission exam, which remained the main entryway to Unicamp, granting access to around 75\% of total slots. By opting for racial quotas, applicants would still write the regular exam but compete for admission only with other eligible quota applicants, which corresponded to between 15\% and 27.2\% of slots depending on the major. These applicants could still opt for PAAIS if they studied in a public school and obtain up to 60 points in each Phase if they completed both secondary and high school in a public institution.\footnote{More precisely, Unicamp granted 40 additional points for high school graduates and 20 points for secondary school graduates in Phases 1 and 2. Unicamp discontinued the racial bonus after implementing the racial quotas.} Second, eligible applicants could alternatively apply by sending their ENEM scores directly to Unicamp to compete for 20\% of the total slots. Under this admission process, Unicamp reserved 50\% of slots for applicants who studied in a public high school, 25\% for Black applicants, and 25\% for Black applicants who graduated from public schools.\footnote{ In 2019, Unicamp also created an admission exam exclusively for native applicants and granted admission to some applicants who received awards in knowledge national and international Olympiads. However, these two processes combined could fill only up to around 5\% of slots. In case of remaining vacancies in these two processes or ENEM, Unicamp transferred the slots back to the regular admission exam.} 


The academic year in Brazil runs from February-March through December. Therefore the academic and calendar year are the same. The admission process at Unicamp uses the incoming year, where admitted students will start their first year in college. The admission process runs from October to January. Students in the final year of high school in 2019 enrolled in the 2020 admission process, which took place between October 2019 and January 2020. Therefore, the onset of the Covid pandemic in March 2020 did not affect the 2020 admission process. The first admission process to be affected by Covid was the 2021 process.

The Covid pandemic has impacted the 2021 and 2022 admission processes in several ways. First, the 2020 ENEM exam took place only in January 2021, instead of November 2020, with the results being released only by the end of March 2021. Since the 2021 academic year started on March 15, Unicamp did not open the ENEM admission process in 2021. Instead, Unicamp filled the ENEM slots with applicants from the regular admission exam following the eligibility criteria (i.e., quotas) defined in the ENEM process. Second, most Brazilian high schools closed nearly during the 2020 and a significant part of the 2021 academic years. During that period, most schools operated remotely with varying degrees of success. We expect this channel to be more pervasive for applicants in public schools during the Covid pandemic.\footnote{At Unicamp's admission exam, around 40\% of applicants recently graduated from high school.} Third, Unicamp and ENEM exams continued to be held in person during the pandemic, possibly reducing the participation of some groups due to the contamination risk.

It is unclear whether the ENEM process suspension significantly impacted application and admission profiles. Indeed, most ENEM applicants also apply through Unicamp's admission exam. For example, in 2019 and 2020, 83\% of ENEM applicants also wrote the Unicamp admission exam. Moreover, Unicamp announced the ENEM suspension in July 2020, before the admission process registration. Therefore, applicants had enough time to register and prepare for the 2021 admission exam, which took place in January 2021 (Phase 1) instead of September 2020.


\section{Data} \label{section:data}

Our main dataset comes from Comvest, Unicamp's admission office.\footnote{\textit{Comissão Permanente para os Vestibulares}, \href{URL}{https://www.comvest.unicamp.br}.} We use data from 2016 to 2022 on all applicants to Unicamp's admission processes, including the regular admission exam, ENEM, indigenous exam, and the Olympiad.\footnote{Our dataset contains an anonymous unique applicant identifier, allowing us to identify the same individual in different admission processes.} The data contain a rich array of applicants' socioeconomic characteristics such as year of birth, gender, race, and self-reported parental background information, including education, occupation, and income. We also observe applicants' major choices and performance in all exams, including ENEM, and whether they were admitted and enrolled in one of Unicamp's majors. Importantly, we also have applicants' names of high schools and graduation years.

To gather information on applicants' high school characteristics, we use public data from the Brazilian School Census and ENEM, the national end-of-high school exam, at the school level for 2015-2019. The School Census reports the type of high school (i.e., public or private) and the total number of students by grade. The ENEM dataset includes the number of test takers by school and applicants' scores.

\subsection{Quality measure}

We first compute a high school quality measure using ENEM scores. For each Brazilian school, we compute the average performance of all students that took the exam between 2015 and 2019. To have a reliable measure, we only consider schools where at least 30 students completed the exam in this time frame. More specifically, we compute the quality measure $q$ for school $j$ as the average score for all students that took the exam in each ENEM subject $k$ between 2015 and 2019.\footnote{ The ENEM exam includes 45 multiple-choice questions on each subject (Language, Mathematics, Science, and Humanities) and an Essay.}
\begin{equation} \label{eq:quality}
q_j = \frac15\sum_{k=1}^5\frac1{N_{j,k}}\sum_{i\in j}\text{score}_{ik}
\end{equation}
We standardize the measure $q$ to have a mean of zero and a standard deviation of one. It is worth noting that we keep the quality measure fixed for all exercises and do not use any data affected by Covid.

\begin{figure}[!h]
    \centering
    \caption{High-school Quality Measure - Brazil}
    \label{fig:quality_density}
    \includegraphics{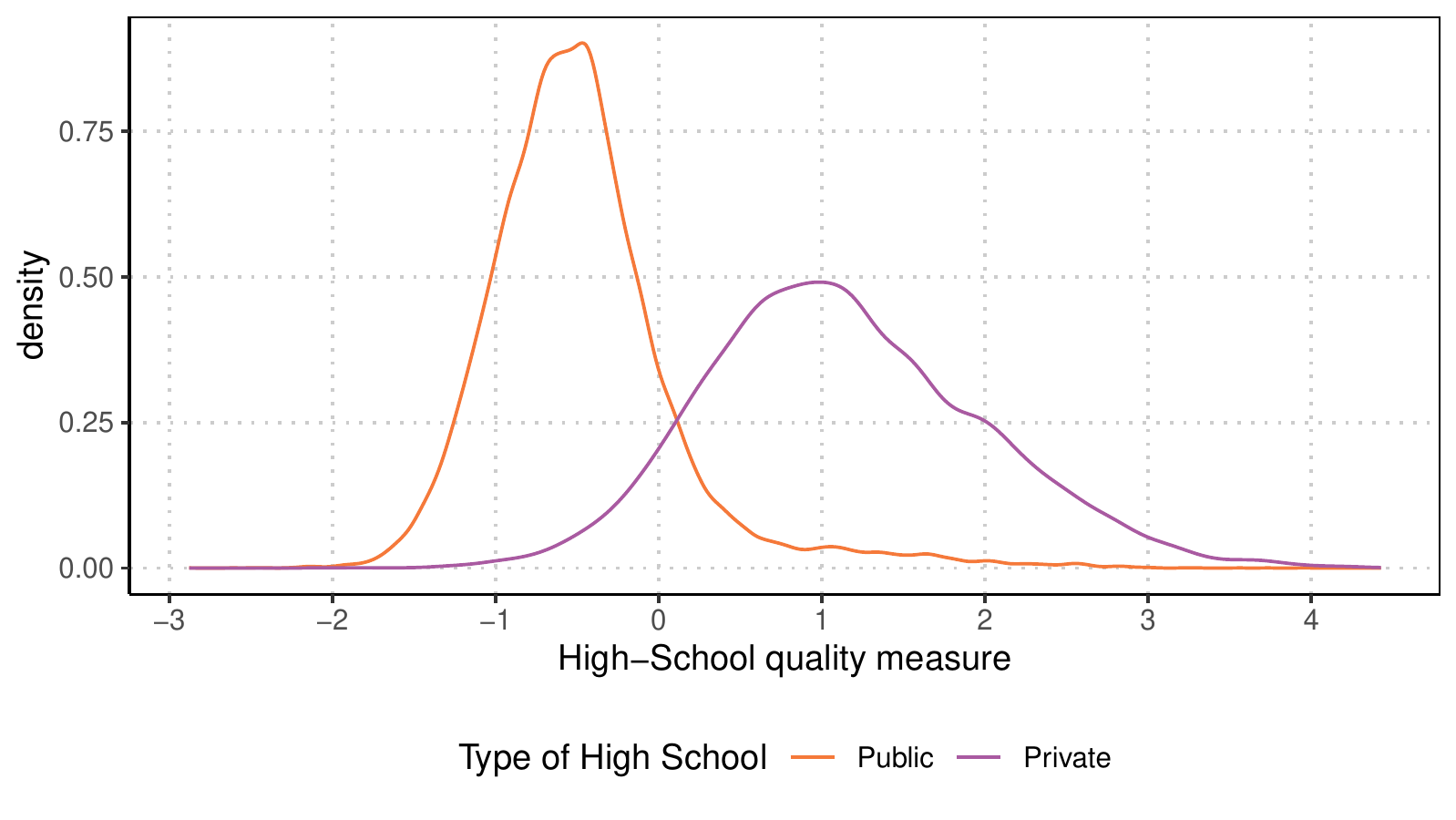}
    \caption*{\footnotesize{Notes: The figures plots the distribution of the quality measure separately for public high schools (in \textcolor{orange}{orange}) and for private schools (in \textcolor{purple}{purple}) in Brazil. We compute the quality measure using the school average in the ENEM exam between 2015-2019 and standardize it to have a mean zero and unitary standard deviation. We include only schools with at least 30 students with valid scores in 2015-2019.}}
\end{figure}

\begin{figure}[!h]
    \centering
    \caption{Proportion of Public High Schools in the Quality Percentiles}
    \label{fig:quality_prop_public}
    \includegraphics{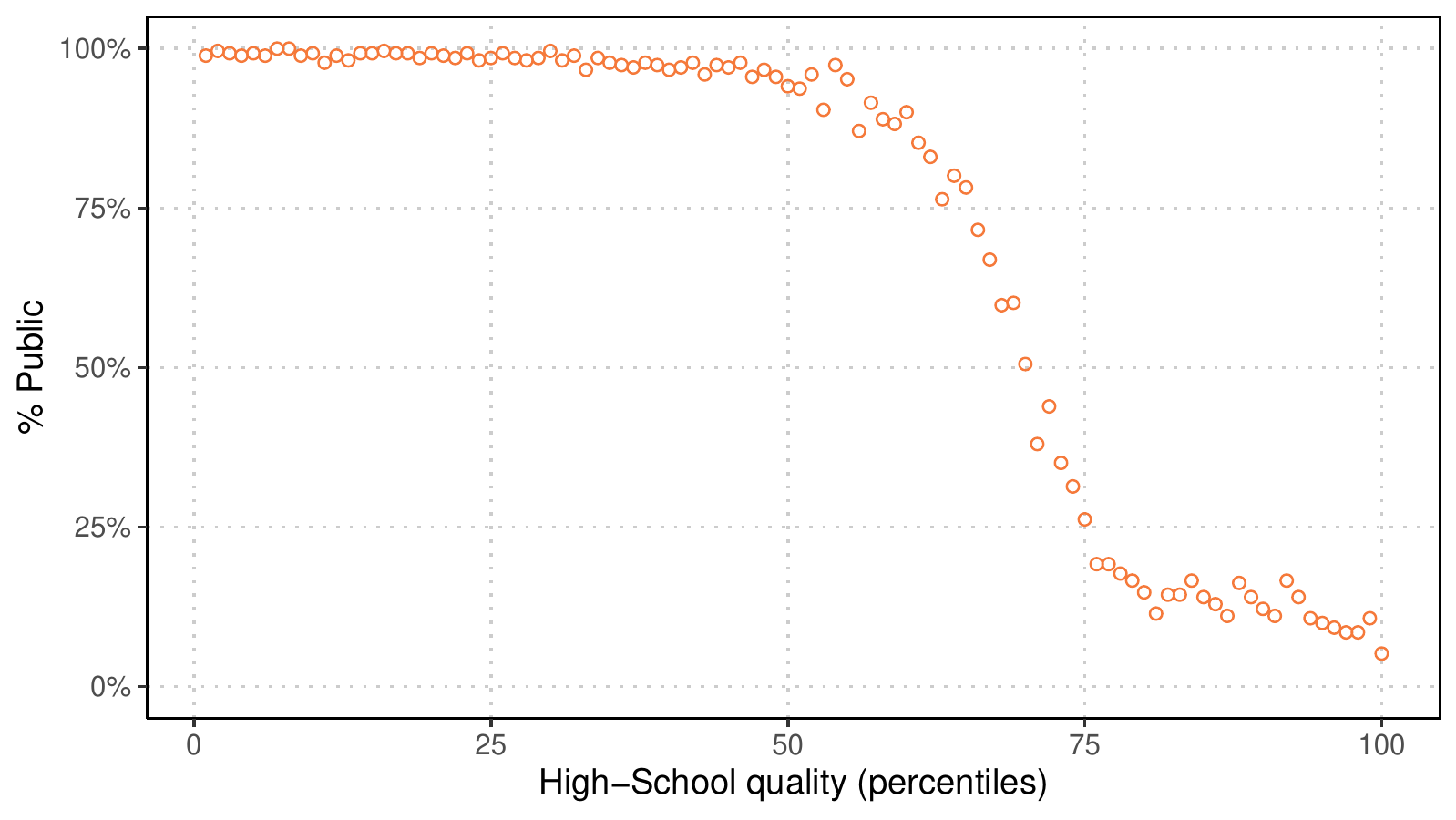}
    \caption*{\footnotesize{Notes: The figure shows the proportion of public high schools along the quality distribution. We divide the quality measure into 100 bins, representing each centile of the distribution.}}
\end{figure}

Figure \ref{fig:quality_density} plots the density of this quality measure for all public and private high schools in the country. It confirms that private schools outperform public schools. The average private school has a quality measure 1.6 standard deviations above the average public school. Figure \ref{fig:quality_prop_public} shows the proportion of public schools for each centile of the quality distribution. We can see that the bottom 50\% schools are predominantly public. Among the top 25\%, only 13.6\% of them are public schools. 

\subsection{Descriptive Statistics}

\begin{table}[!h]
\centering
\caption{Descriptive Statistics}\label{tab:desc_stat}
\begin{adjustbox}{max width = \textwidth, width = \textwidth, center}
\begin{threeparttable}
\begin{tabular}[t]{lrrrcrrr}
\toprule
& \multicolumn{3}{c}{Applicants} & & \multicolumn{3}{c}{Enrolled} \\ \cmidrule{2-4} \cmidrule{6-8}
 Variable & \multicolumn{1}{c}{Mean} & \multicolumn{1}{c}{Std-Dev} & \multicolumn{1}{c}{\# Obs} & & \multicolumn{1}{c}{Mean} & \multicolumn{1}{c}{Std-Dev} & \multicolumn{1}{c}{\# Obs} \\
 \hline \addlinespace
 Age & 20.087 & 3.159 & \num{513144} && 20.487 & 4.063 & \num{23243}\\ 
 \addlinespace
  Black/Native & 0.225 & 0.418 & \num{502525} && 0.283 & 0.450 & \num{22650}\\
 \addlinespace
 Mother w/ less than High-School degree & 0.159 & 0.365 & \num{507622} && 0.157 & 0.364 & \num{22962}\\
 \addlinespace
 Income less 3 minimum wages & 0.286 & 0.452 & \num{507622} && 0.252 & 0.434 & \num{22962}\\
 \addlinespace 
 Directly from High-School & 0.445 & 0.497 & \num{513144} && 0.396 & 0.489 & \num{23243}\\
 \addlinespace 
 Has High-School information & 0.931 & 0.253 & \num{513144} && 0.944 & 0.229 & \num{23243} \\ 
 \addlinespace
 Private High-School & 0.624 & 0.484 & \num{434037} && 0.509 & 0.500 & \num{20049} \\ 
 \addlinespace 
 High School quality (standardized) & 1.404 & 1.166 & \num{431528} && 1.613 & 1.106 & \num{19984} \\
 \addlinespace 
 High School quality percentile & 82.047 & 19.878 & \num{431528} &&
 85.423& 17.986 & \num{19984} \\
 \addlinespace
 Enrolled & 0.045 & 0.208 & \num{513144} && 1.000 & 0.000 & \num{23243} \\
\bottomrule
\end{tabular}
\begin{tablenotes}
    \item \footnotesize{Notes: The table shows each variable's mean, standard deviation, and the number of valid observations. Each row represents a unique variable, indicated in the first column. The next three columns show the information for all applicants, and the last three are for the sample of enrolled students. The table uses information for the entire period between 2016 and 2022.} 
\end{tablenotes}
\end{threeparttable}
\end{adjustbox}
\end{table}

Table \ref{tab:desc_stat} shows descriptive statistics from our Unicamp data, including each applicant's high-school quality measure  calculated using equation (\ref{eq:quality}). We show the mean, standard deviation, and the number of valid observations for each variable, separately for the entire sample of applicants and the sub-sample of enrolled students.

The average Unicamp applicant is 20 years old, around 22.5\% are Black or Native, 15.9\% have mothers with less than a high-school degree, and 28.6\% have a family income lower than three minimum wages. On average, 44.5\% of students apply directly from high school, i.e., there is no gap year between their high school graduation and college application. 

When we combine the Unicamp dataset with the School Census and ENEM datasets, we obtain school information for 93.1\% of the applicants' sample, for which we can get the type of schools (private or public) and their quality scores. The positive selection in the pool of applicants and enrollees is clear. The average applicant comes from a school with a quality 1.40 standard deviations above average, positioned at the 82nd percentile. The average enrolled candidate is even more positively selected from schools with 1.61 standard deviations above average and in the 85th percentile. Only 4.5\% of applicants enroll at the university, which is not surprising, given how selective is the admission process. 

\section{Results} \label{section:results}

Our analysis focuses on three periods around the Covid pandemic. The first is the pre-quota period (2016-2018) when only the bonus version of Unicamp's affirmative action policy was in place. The second is the quota period (2019-2020), in which Unicamp reserved a significant proportion of slots for public high school graduates and racial minorities. Finally, we pay particular attention to the admission years impacted by the Covid pandemic (2021-2022) when the quota policy was still in place.\footnote{ Section \ref{section:institutional} contains more details on Unicamp's admission policies and how the Covid pandemic may have affected them. As explained, the 2020 admission process took place by the end of 2019, not affected by the pandemic.}

\subsection*{Fact 1: Admission is concentrated at the top of the distribution}

\begin{figure}[!h]
    \caption{High School Quality and Unicamp Application and Enrollment}\label{fig:pre_raw}
    \includegraphics[width=\textwidth]{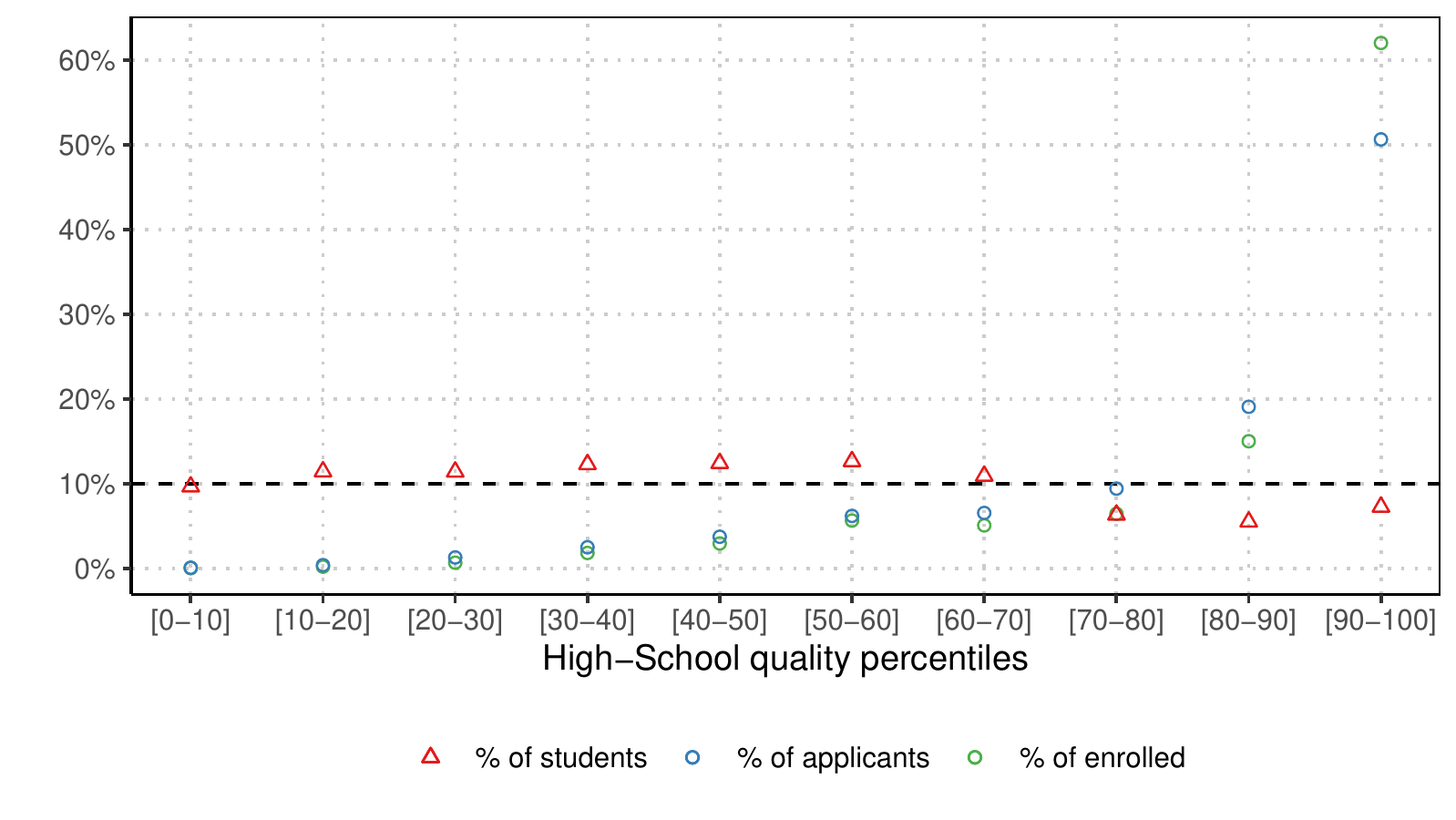}
    \caption*{\footnotesize{Notes: The figure plots the percentage of students enrolled in high school (\textcolor{red}{red} triangles) for schools in each decile of the quality distribution. We also compute what fraction of applicants at Unicamp belonged to schools in each decile (in \textcolor{blue}{blue} circles). Similarly, we present the fraction of enrolled students by school decile (in \textcolor{green}{green} circles).}}
\end{figure}

We first show that admission to a selective university like Unicamp is concentrated at the top of the high school quality distribution, using data from the pre-quota period (2016-2018). Figure \ref{fig:pre_raw} shows for each decile of the quality distribution the proportion of students enrolled in those schools (in \textcolor{red}{red}), the proportion of students applying to Unicamp (in \textcolor{blue}{blue}), and the proportion of students enrolled at Unicamp (in \textcolor{green}{green}) in 2016-2018. The proportion of students applying and enrolling at Unicamp is increasing in the high school quality percentiles. 

Moreover, there is a sharp discontinuity in the association between application/enrollment and high school quality as we move to the top of the quality distribution. In particular, the top 10\% schools enroll less than 10\% of students in the country but account for 50\% and 62\% of Unicamp applicants and enrollees, respectively.

Another way of detecting this concentration is by focusing on the top 100 private schools (out of 8,962 private schools in the country) and the top 100 public schools (out of 20,596 public schools in the country). Together, they correspond to 0.68\% of schools and 15\% of students who enrolled at Unicamp. The concentration at the top seems particularly pervasive in the public school system. Breaking this number by type of school, among the admitted students from private high schools, 8\% were in the top 100. Among admitted public students, 21.4\% were in the top 100 public high schools.

While we would naturally expect students from higher-quality schools to have better performance, the high concentration of students coming from a handful of schools seem problematic. Low SES students are significantly less likely to attend these high-quality schools. With individual talent and ability being dispersed over the school distribution, Unicamp was likely missing high-achieving students not enrolled in top schools. This has also been documented in other contexts by \cite{hoxby13} and \cite{dynarski21}. 
 
\subsection*{Fact 2: Introduction of quotas for public high school students and racial minorities reduced the concentration at the top}

We now show how introducing quotas in 2019 reduced the concentration of enrolled students from top schools. To do so, we run the following regression for the 2016-2020 admission years:
\begin{equation} \label{eq:quotas}
Y_{j(i)} = \alpha + \beta \text{Quota}_{t(i)} + \varepsilon_{i},
\end{equation}
\noindent where $Y_{j(i)}$ is a characteristic for the school $j$ where enrolled student $i$ obtained their high school degree and $\text{Quota}_{t(i)}$ is a dummy variable equal to one if quotas were in place in time $t$ when individual $i$ applied. We clustered the standard errors at the individual level, taking into account that the same individual can apply in different years. 

\begin{figure}[!h]
    \centering
    \caption{Quota and High-School Concentration}
    \label{fig:effects_quota}
    \includegraphics{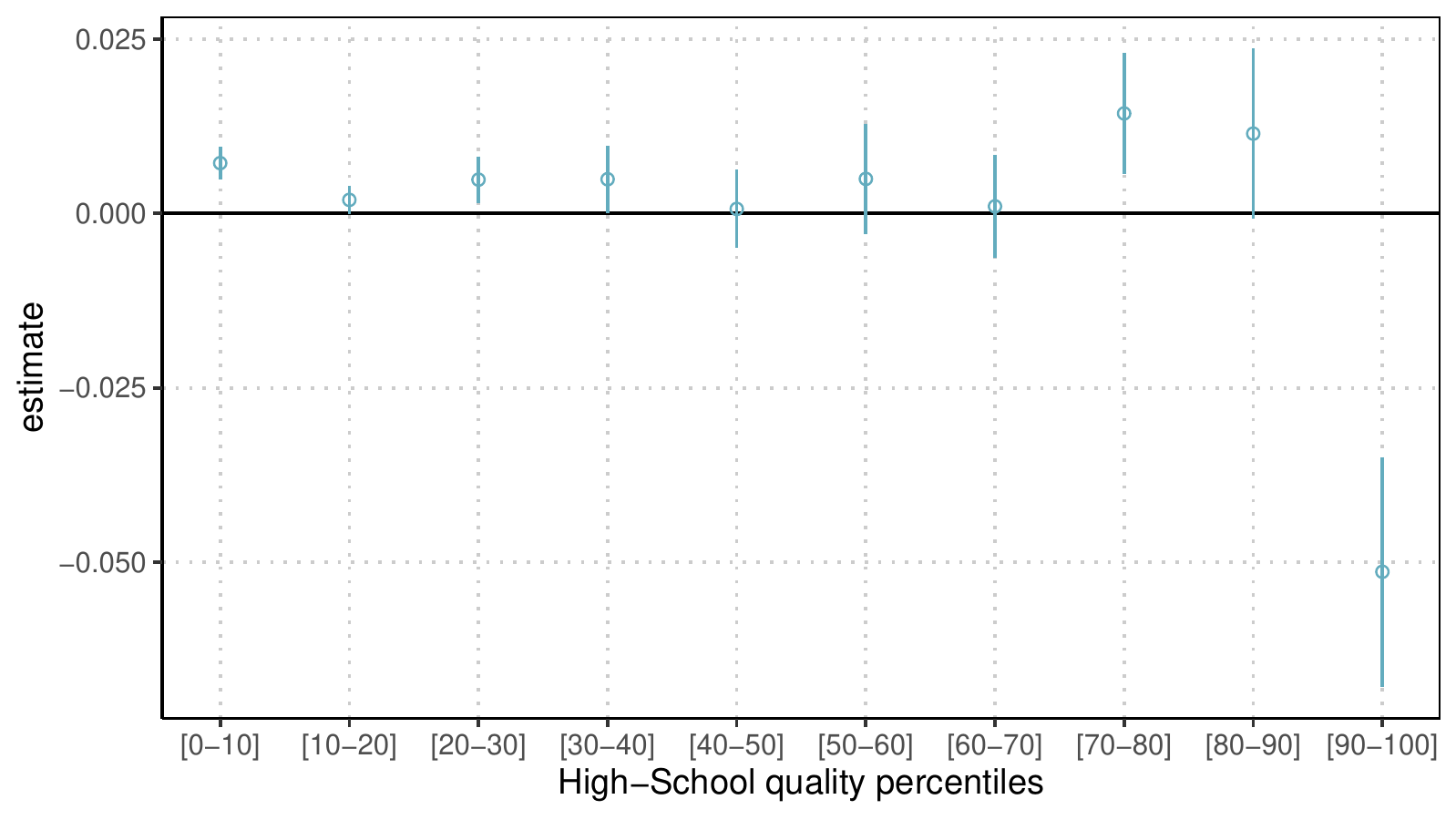}
    \caption*{\footnotesize{Notes: The figure shows the $\beta$ coefficients from equation (\ref{eq:quotas}). Each dot represents the coefficient for a different regression. The outcome variable is whether the enrolled student graduated from a high school in each of the ten deciles of the quality distribution. The standard errors are clustered at the individual level. The figure also displays the 95\% confidence intervals.}} 
\end{figure}

Figure \ref{fig:effects_quota} shows the $\beta$ coefficients for this regression, where the outcome variable is whether school $j$ belongs to each decile of the quality distribution. After introducing the quota policy, the probability of enrolled students coming from the top 10\% schools is reduced by five percentage points. The beneficiaries are schools in the 70-90 percentiles and those below 40. It is worth noting that, for schools in the bottom 40\%, while the magnitude of the estimates does not seem large, the proportion of admitted students from these schools is extremely small. So, this increase represents a large gain relative to the baseline period.

 Table \ref{tab:effects_quota} shows the intercept and $\beta$ coefficients for different quality-related outcomes. In the first column, we see that while in the baseline period, 15\% of admitted students came from the top-100 private and public schools, this is reduced by 2.7 percentage points with the introduction of quotas. In the second column, we show that the quota is associated with a reduction of 0.10 standard deviations in the average quality of enrolees' schools of origin. In the third column, we can see that the average student came from a school in the 86th percentile in 2016-2018, but 2.1 percentiles less after the quota implementation. Finally, in the last four columns, we verify that the shift away from the top 10\% (reduction of 5.1pp) benefitted students from schools in the top 70-90 (2.6pp) and the bottom 40 (1.9pp). These gains are particularly large for the schools in the bottom 40, representing an almost 70\% increase relative to the baseline.

\begin{table}[!h]
\centering
\caption{Quotas and Enrollees' High School Background}\label{tab:effects_quota}
\begin{threeparttable}
\begin{tabular}[t]{lccccccc}
\toprule
 && & \multicolumn{5}{c}{Quality Percentiles}\\ \cmidrule(l{10pt}r{10pt}){4-8}
\addlinespace
 Outcome &  Top-100 & Quality& Avg & [0-40] & [40-70] & [70-90] & [90-100] \\ 
 \addlinespace
  & (1) & (2) & (3) & (4) & (5) & (6) & (7) \\ 
\midrule
(Intercept) & \num{0.150} & \num{1.621} & \num{85.916} & \num{0.028} & \num{0.137} & \num{0.215} & \num{0.620}\\
 & (\num{0.004}) & (\num{0.012}) & (\num{0.189}) & (\num{0.002}) & (\num{0.004}) & (\num{0.005}) & (\num{0.005})\\
 \\
Quota & \num{-0.027} & \num{-0.095} & \num{-2.093} & \num{0.019} & \num{0.007} & \num{0.026} & \num{-0.051}\\
 & (\num{0.006}) & (\num{0.019}) & (\num{0.317}) & (\num{0.003}) & (\num{0.006}) & (\num{0.007}) & (\num{0.008})\\
\midrule
Num.Obs. & \num{14158} & \num{14158} & \num{14158} & \num{14158} & \num{14158} & \num{14158} & \num{14158}\\

\bottomrule
\end{tabular}
\begin{tablenotes}
    \item \footnotesize{Notes: The table shows the intercept and $\beta$ coefficients of the quota dummy from equation (\ref{eq:quotas}) for different outcome variables. In the first column, the dependent variable is a dummy variable equal to one if the enrolled student graduated from a top 100 public or top 100 private high schools and zero otherwise. In the second column, the outcome variable is the quality measure defined by equation (\ref{eq:quality}) standardized to have a mean of zero and a standard deviation of one. In the third column, the dependent variable is the quality percentile of each school. In the last four columns, the outcomes are dummy variables equal to one if the high school of origin of the enrolled student belongs to each quality percentile range and zero otherwise. We cluster the standard errors at the individual level.} 
\end{tablenotes}
\end{threeparttable}
\end{table}

 \subsection*{Fact 3: Covid reduced the gains from the quota policy by increasing concentration at the top} 

We now rerun the exercises above to assess the impact of Covid on high school quality concentration among Unicamp enrollees by including data on the 2021-2022 admission processes. To do so, we augment the equation (\ref{eq:quotas}) by adding a dummy variable equal to one if the student $i$ was admitted during the Covid pandemic (i.e., in 2021 or 2022) and zero otherwise.

\begin{figure}[!h]
    \centering
    \caption{Covid and High-School Concentration}
    \label{fig:effects_covid}
    \includegraphics{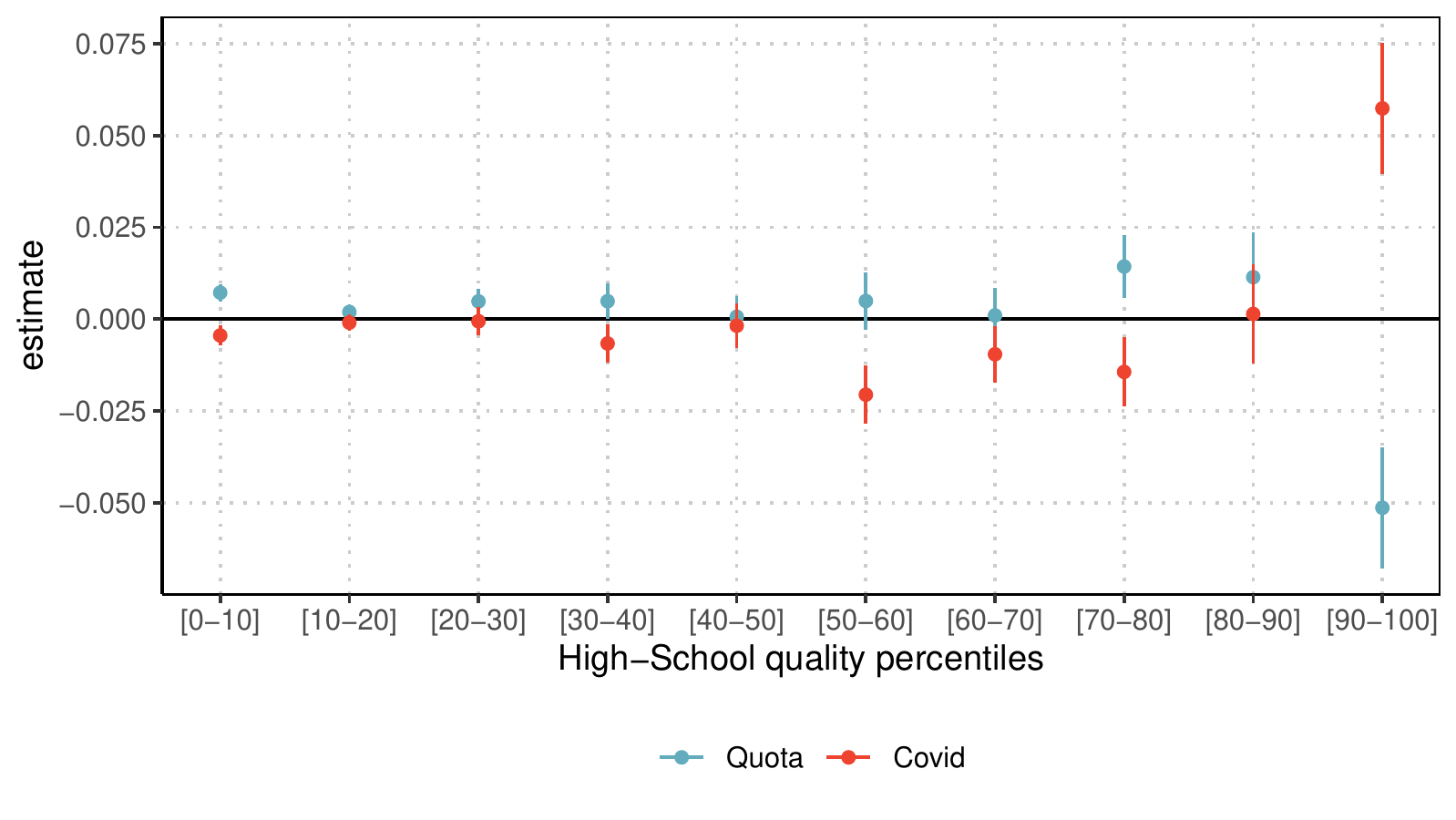}
    \caption*{\footnotesize{Notes: The figure shows the $\beta$ coefficients from the augmented version of equation (\ref{eq:quotas}) that includes quota and Covid dummies. Each pair of dots represents the coefficients for a different regression. The outcome variable is whether the enrolled student graduated from a high school in each of the ten deciles of the quality distribution. The standard errors are clustered at the individual level. The figure also displays the 95\% confidence intervals.}} 
\end{figure}

Figure \ref{fig:effects_covid} presents the estimated coefficients for the quota (i.e., 2019-2022) and Covid (i.e., 2021-2022) dummies, having as a dependent variable whether the school $j$ where the student $i$ graduated from belongs to each decile of the quality distribution. The Covid pandemic's impact on enrolled students' school background is the mirror image of the quota, nearly offsetting the original impact of the quota policy. Students from high schools at the top 10\% of the quality distribution experienced large gains in their probability of enrollment, mostly at the expense of students who graduated from high schools located in the 50-80 deciles. 

To better quantify the magnitudes of the impact of the Covid pandemic, we present regression results in Table \ref{tab:effects_covid} for different high school quality outcome measures. All measures point to a concentration increase at top-quality high schools, in contrast to what was achieved by the quota policy. 

Indeed, considering their high school quality background, the Covid pandemic had substantial distributive effects on enrolled students' profiles. Covid increased concentration at the top-100 public and private schools by an amount representing 70\% of the reduction in concentration achieved by the quota policy. In terms of the average quality of schools of origin, the Covid pandemic completely offset the impact of the quota policy. After Covid, the average student came from a school located in the 86 percentile, as before the quota. Moreover, only schools in the top 10\% of the high school quality distribution experienced gains in terms of the number of students enrolled at Unicamp in 2021-2022. These gains erased the losses due to the quota policy for this group. The opposite happened at the bottom 40\%: Covid completely offset any gains associated with the quota policy for this group. In addition, the average student in a school in the 40-70 deciles, a group that had not previously benefited from the quota policy, experienced a reduction of 23\% in enrollment. Finally, for the 70-90 deciles of the quality distribution, half of the gains previously generated by the quota disappeared in the admission years affected by the Covid pandemic. 

\begin{table}[!h]
\centering
\caption{Covid and Enrollees' High School Background}\label{tab:effects_covid}
\begin{threeparttable}
\begin{tabular}[t]{lccccccc}
\toprule
 && & \multicolumn{5}{c}{Quality Percentiles}\\ \cmidrule(l{10pt}r{10pt}){4-8}
\addlinespace
 Outcome &  Top-100 & Quality& Avg & [0-40] & [40-70] & [70-90] & [90-100] \\ 
 \addlinespace
  & (1) & (2) & (3) & (4) & (5) & (6) & (7) \\ 
\midrule
(Intercept) & \num{0.150} & \num{1.621} & \num{85.916} & \num{0.028} & \num{0.137} & \num{0.215} & \num{0.620}\\
 & (\num{0.004}) & (\num{0.012}) & (\num{0.189}) & (\num{0.002}) & (\num{0.004}) & (\num{0.005}) & (\num{0.005})\\
 \\
Quota & \num{-0.027} & \num{-0.095} & \num{-2.093} & \num{0.019} & \num{0.007} & \num{0.026} & \num{-0.051}\\
 & (\num{0.006}) & (\num{0.019}) & (\num{0.317}) & (\num{0.003}) & (\num{0.006}) & (\num{0.007}) & (\num{0.008})\\
 \\
Covid & \num{0.019} & \num{0.161} & \num{2.455} & \num{-0.013} & \num{-0.032} & \num{-0.013} & \num{0.057}\\
 & (\num{0.006}) & (\num{0.021}) & (\num{0.345}) & (\num{0.004}) & (\num{0.006}) & (\num{0.008}) & (\num{0.009})\\
\midrule
Num.Obs. & \num{19984} & \num{19984} & \num{19984} & \num{19984} & \num{19984} & \num{19984} & \num{19984}\\

\bottomrule
\end{tabular}
\begin{tablenotes}
    \item \footnotesize{Notes: The table shows the intercept and $\beta$ coefficients of the quota and Covid dummies from the augmented version (\ref{eq:quotas}) for different outcome variables. In the first column, the dependent variable is a dummy variable equal to one if the enrolled student graduated from a top 100 public or top 100 private high schools and zero otherwise. In the second column, the outcome variable is the quality measure defined by equation (\ref{eq:quality}) standardized to have a mean of zero and a standard deviation of one. In the third column, the dependent variable is the quality percentile of each school. In the last four columns, the outcomes are dummy variables equal to one if the high school of origin of the enrolled student belongs to each quality percentile range and zero otherwise. We cluster the standard errors at the individual level.} 
\end{tablenotes}
\end{threeparttable}
\end{table}

Our results so far point to heterogeneous impacts of the Covid pandemic depending on applicants' high school quality. In principle, these differential effects could be due to several channels. First, schools may have differed in their capacity to adapt to online teaching, affecting students' learning in the last one or two years of high school. Since the admission exam tests applicants' knowledge of high school disciplines, school interruptions and online classes could have hindered students' ability to perform well in the admission exam, especially in relatively lower-performing schools. Second, schools may also play an important role in helping their students navigate the college admission process. If this process is more difficult to achieve remotely in lower-quality schools, their students may not even apply to university. Third, high school quality may be just a proxy for students' disadvantaged backgrounds, rendering them particularly susceptible to shocks, such as the Covid pandemic.

To gain insight into the mechanisms behind the strong association between high school quality and the impact of the Covid pandemic, we first test whether the effects are driven by students attending high school during the pandemic. Indeed, we expect that the direct impact on learning or support during the college application process should predominantly affect students attending school during the Covid pandemic.

\begin{table}[!h]
\centering
\caption{Covid Effects by High School Status during Covid}\label{tab:covid_schools}
\begin{threeparttable}
\begin{tabular}[t]{lccccccc}
\toprule
 && & \multicolumn{5}{c}{Quality Percentiles}\\ \cmidrule(l{10pt}r{10pt}){4-8}
\addlinespace
 Outcome &  Top-100 & Quality& Avg & [0-40] & [40-70] & [70-90] & [90-100] \\ 
 \addlinespace
  & (1) & (2) & (3) & (4) & (5) & (6) & (7) \\ 
\midrule
Quota & \num{-0.030} & \num{-0.114} & \num{-2.369} & \num{0.020} & \num{0.010} & \num{0.028} & \num{-0.058}\\
 & (\num{0.006}) & (\num{0.019}) & (\num{0.311}) & (\num{0.003}) & (\num{0.006}) & (\num{0.007}) & (\num{0.008})\\
 \\
Not in school during Covid  & \num{0.002} & \num{0.024} & \num{0.148} & \num{0.003}  & \num{-0.008} & \num{-0.007} & \num{0.012}\\
 \hphantom{a}× Covid  & (\num{0.007}) & (\num{0.028}) & (\num{0.507}) & (\num{0.006}) & (\num{0.009}) & (\num{0.011}) & (\num{0.012})\\
 \\ 
In school during Covid& \num{0.021} & \num{0.202} & \num{3.272} & \num{-0.019} & \num{-0.039} & \num{-0.011} & \num{0.070}\\
  \hphantom{a} × Covid & (\num{0.008}) & (\num{0.024}) & (\num{0.348}) & (\num{0.003}) & (\num{0.007}) & (\num{0.009}) & (\num{0.011})\\
\midrule
Num.Obs. & \num{19984} & \num{19984} & \num{19984} & \num{19984} & \num{19984} & \num{19984} & \num{19984}\\

\bottomrule
\end{tabular}
\begin{tablenotes}
    \item \footnotesize{Notes: The table shows the coefficients associated with a dummy for quota and an interaction between the dummy for Covid and a binary variable for whether the student was in high-school during Covid (2020 or 2021). In the first column, the dependent variable is a dummy variable equal to one if the enrolled student graduated from a top 100 public or top 100 private high school and zero otherwise. In the second column, the outcome variable is the quality measure defined by equation (\ref{eq:quality}) standardized to have a mean of zero and a standard deviation of one. In the third column, the dependent variable is the quality percentile of each school. In the last four columns, the outcomes are dummy variables equal to one if the high school of origin of the enrolled student belongs to each quality percentile range and zero otherwise. We cluster the standard errors at the individual level.} 
\end{tablenotes}
\end{threeparttable}
\end{table}

Table \ref{tab:covid_schools} presents the results of our main specification by interacting the Covid dummy with a dummy variable equal to one if the applicant was in high school during the 2020 and/or 2021 school year.\footnote{This specification also includes a binary control for whether the student graduated from high school in the same year. Therefore, the interaction terms capture only the effect of studying during Covid, not whether the student was a recent high school graduate.} The results show that the differentiated effect by high school quality seems to be driven exclusively by applicants hit by the Covid pandemic during high school. Indeed, all coefficients for those graduating before the Covid pandemic are indistinguishable from zero. Thus, while we cannot rule out the impact of students' backgrounds, the results in Table \ref{tab:covid_schools} suggest that the school shutdowns and online classes directly interfered with high schools' role in the transition to college.

\begin{figure}[!h]
    \centering
    \caption{Covid Effects on Applicants and Enrolled Students}
    \label{fig:effects_covid_applicants}
    \includegraphics{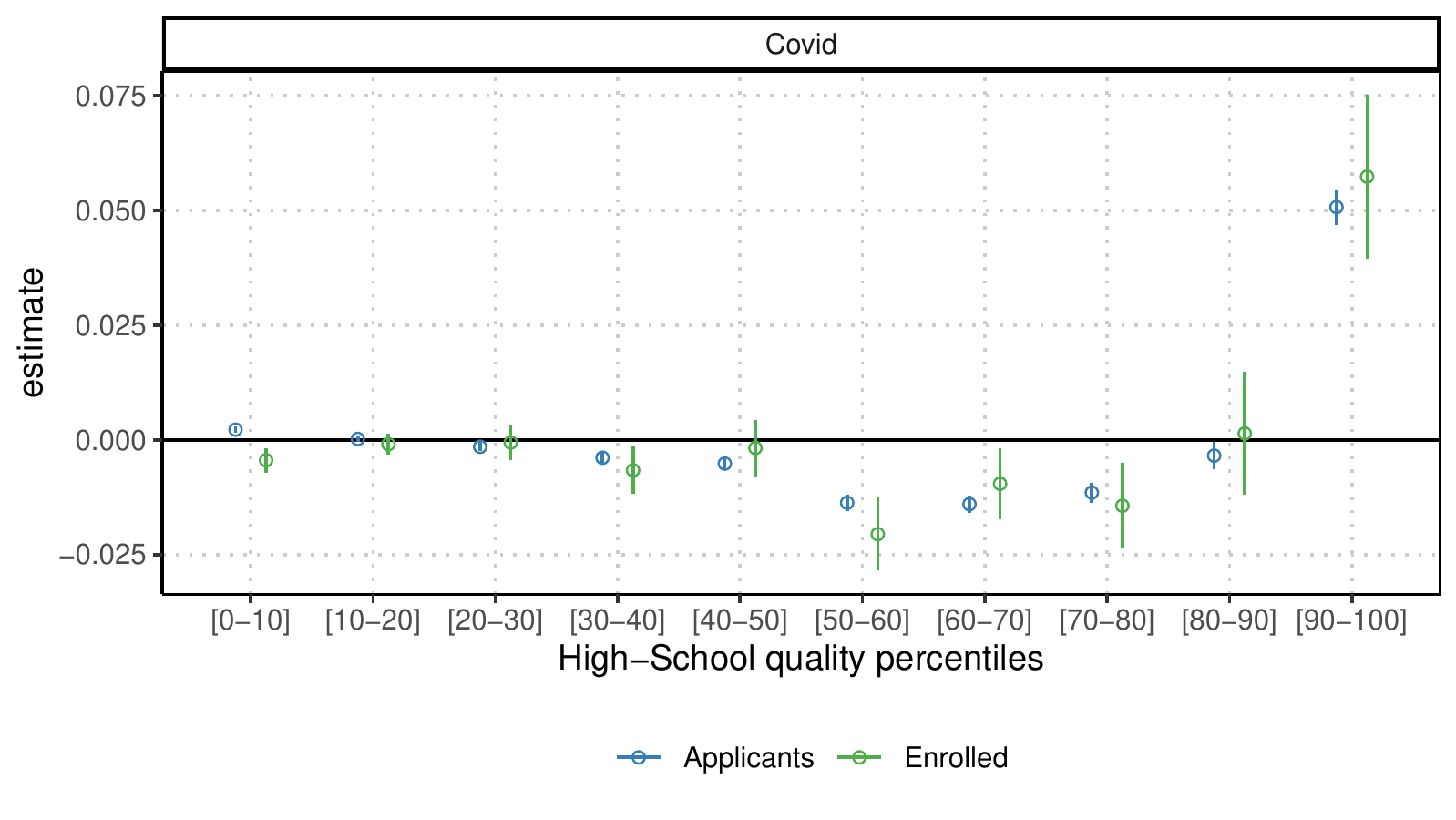}
    \caption*{\footnotesize{Notes: The figure shows the $\beta$ coefficients for the binary Covid variable from the augmented version of the equation (\ref{eq:quotas}) that includes quota and Covid dummies. Each pair of dots represents the coefficients for a different regression on a different outcome. The outcome variable is whether the applicant graduated from a high school in each of the ten deciles of the quality distribution (in \textcolor{blue}{blue}) or the equivalent, restricted for the enrolled students (in \textcolor{green}{green}). The standard errors are clustered at the individual level. The figure also displays the 95\% confidence intervals.}} 
\end{figure}

To further investigate whether learning or other aspects of the college application process were mostly affected by Covid, we investigate whether students' application behavior accounts for the observed impacts on Unicamp enrollment. Figure \ref{fig:effects_covid_applicants} presents the estimated coefficients for Covid, considering application and enrollment as two separate outcome variables. Comparing both outcomes suggests that they tend to move close to each other with Covid, so part of the losses experienced by lower-achieving schools may be due to fewer applications from their students. Still, we cannot rule out that learning deficits due to the Covid shock may partially explain these impacts, as they can also affect the likelihood of applying to Unicamp.

\clearpage
\subsection{Robustness} 

We now present additional exercises to show that our results are not driven by time trends, specific choices related to our quality measure, or that the Covid pandemic may have disproportionally hit out-of-state applicants.

\begin{figure}[!h]
    \centering
    \caption{Average High School Quality Percentile by Year}
    \label{fig:effects_yearly}
    \includegraphics{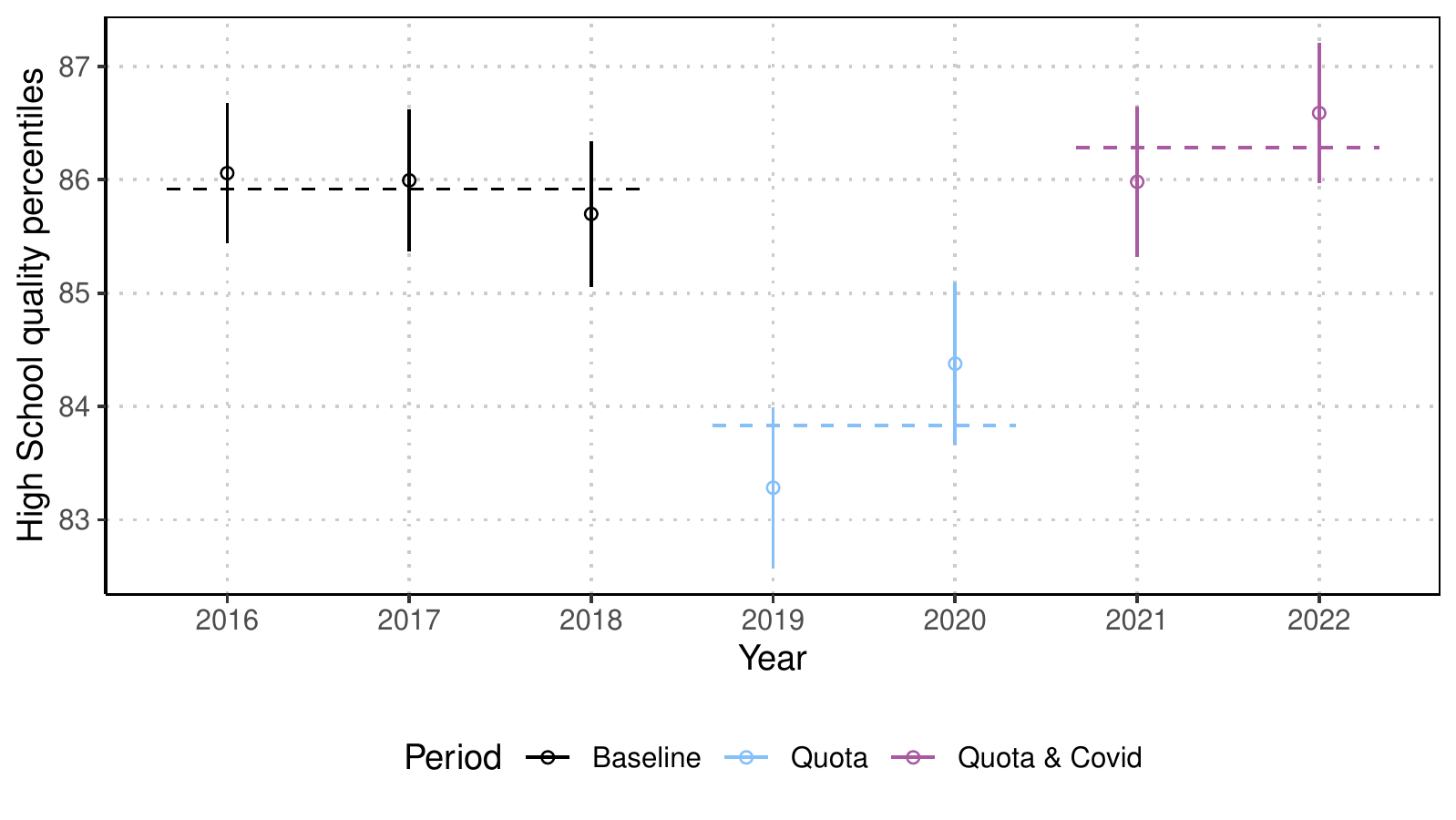}
    \caption*{\footnotesize{Notes: The figure shows the average high school quality percentile for all students enrolled at Unicamp each year from 2016 to 2022. The colors represent the three distinct periods analyzed in the paper, the baseline (2016-2018, in \textcolor{black}{black}), the period with only quota (2019-2020, in \textcolor{blue}{blue}), and the admission years with quota and Covid (2021-2022, in \textcolor{purple}{purple}). The dashed lines represent the mean for each of the three periods. The solid lines for each year are the 95\% confidence interval.}} 
\end{figure}

To address the possibility that we may be capturing time trends, we compute the yearly average high school quality percentile, one of our outcome variables. Figure \ref{fig:effects_yearly} shows that the average high school quality was similar in the years before the quota implementation. In 2019, there was a substantial decline in the average high school quality percentile of Unicamp enrollees. While there is an increase in 2020, the confidence intervals of 2019 and 2020 overlap. The Covid pandemic, which affected the 2021 and 2022 admission processes, has boosted the average high school quality percentile of enrollees, kept at a high level both in 2021 and 2022.
 
In our main analysis, our high school quality measure is unconditional, meaning that we use the overall school distribution to compute the quality percentiles. However, public and private schools are unequally distributed in the quality percentiles, as shown in Figure \ref{fig:quality_prop_public}. Since they may also be differentially affected by the quota policy and the Covid pandemic, a potential concern is that a conditional quality percentile measure could lead to different conclusions. 

Table \ref{tab:effects_covid_conditional} presents the impact of the quota policy and the Covid pandemic on the conditional high school quality percentiles of enrollees. While the magnitudes of certain coefficients change, our overall conclusions remain unchanged. Indeed, the Covid pandemic led to an expressive redistribution of enrollees toward the top 10\% high schools at the expense of students from high schools located elsewhere in the quality distribution.

\begin{table}[!h]
\centering
\caption{Covid and Enrollees’ High School Background (conditional) }\label{tab:effects_covid_conditional}
\begin{threeparttable}
\begin{tabular}[t]{lccccc}
\toprule
 &\multicolumn{5}{c}{Quality Percentiles (conditional)}\\ \cmidrule(l{10pt}r{10pt}){2-6}
\addlinespace
 Outcome: & Avg & [0-40] & [40-70] & [70-90] & [90-100] \\ 
 \addlinespace
  & (1) & (2) & (3) & (4) & (5)  \\ 
\midrule
(Intercept) & \num{85.453} & \num{0.041} & \num{0.155} & \num{0.207} & \num{0.585}\\
 & (\num{0.206}) & (\num{0.002}) & (\num{0.004}) & (\num{0.004}) & (\num{0.005})\\
 \\
Quota & \num{-2.939} & \num{0.025} & \num{0.012} & \num{0.009} & \num{-0.046}\\
 & (\num{0.348}) & (\num{0.004}) & (\num{0.006}) & (\num{0.007}) & (\num{0.008})\\
 \\
Covid & \num{1.853} & \num{-0.017} & \num{-0.041} & \num{-0.027} & \num{0.036}\\
 & (\num{0.385}) & (\num{0.004}) & (\num{0.007}) & (\num{0.007}) & (\num{0.009})\\
\midrule
Num.Obs. & \num{19984} & \num{19984} & \num{19984} & \num{19984} & \num{19984}\\

\bottomrule
\end{tabular}
\begin{tablenotes}
    \item \footnotesize{Notes: The table shows the intercept and $\beta$ coefficients of the quota and Covid dummies from the augmented version (\ref{eq:quotas}) for different outcome variables. In all columns the quality measure percentile is computed separately for public and private schools. In the first column dependent variable is the quality percentile of each school (conditional on public/private). In the last four columns, the outcomes are dummy variables equal to one if the high school of origin of the enrolled student belongs to each quality percentile range and zero otherwise. We cluster the standard errors at the individual level.} 
\end{tablenotes}
\end{threeparttable}
\end{table}

Another potential concern is that our high school quality measure, calculated using schools' ENEM scores, may not capture relevant school characteristics that make them more vulnerable to the Covid shock. Instead of relying on that measure, we can use the socioeconomic demographic characteristics of Unicamp applicants and check how their enrollment prospects were affected by the quota and Covid.

Table \ref{tab:covid_stu_characteristics} presents results considering enrollees' racial background, mother's education level, and low-income status as outcome variables. These results confirm qualitatively our previously detected effects: the quota policy increased enrollment of relatively less advantaged groups, and the Covid pandemic acted in the opposite direction by reducing their likelihood to enroll at Unicamp, even in the presence of quotas.

\begin{table}
\centering
\caption{Covid Effects and Students Characteristics}\label{tab:covid_stu_characteristics}
\begin{threeparttable}
\begin{tabular}[t]{lccc}
\toprule
 Outcome &  Black-Native & Mother Less High-School & Low-Income \\ 
 \addlinespace
  & (1) & (2) & (3)\\ 
\midrule
(Intercept) & \num{0.234} & \num{0.159} & \num{0.211}\\
 & (\num{0.004}) & (\num{0.004}) & (\num{0.004})\\
 \\
Quota & \num{0.102} & \num{0.017} & \num{0.079}\\
 & (\num{0.007}) & (\num{0.006}) & (\num{0.007})\\
 \\
Covid & \num{-0.040} & \num{-0.043} & \num{-0.019}\\
 & (\num{0.008}) & (\num{0.006}) & (\num{0.008})\\
\midrule
Num.Obs. & \num{22650} & \num{22878} & \num{22962}\\
\bottomrule
\end{tabular}
\begin{tablenotes}
    \item \footnotesize{Notes: The table shows the intercept and $\beta$ coefficients of the quota and Covid dummies from the augmented version (\ref{eq:quotas}) for different outcome variables. In the first column, the dependent variable is a dummy variable equal to one if the student self-reported as Black or Native. In the second column, the dependent variable is a dummy for whether the student's mother has less than a high school degree. In the last column, the outcome is a binary variable for whether the student is low-income. We cluster the standard errors at the individual level.} 
\end{tablenotes}
\end{threeparttable}
\end{table}

Finally, the previous analysis included applicants from Sao Paulo, where 85\% of Unicamp students originate, and out-of-state applicants. A potential concern is that the Covid pandemic may have affected out-of-state applicants disproportionately, as Unicamp's admission exam was held in person, and Unicamp canceled the ENEM process in 2021. To address that concern, we rerun the analysis considering only applicants residing in the state of Sao Paulo. Table \ref{tab:effects_SP} shows results similar to our main specification.

\begin{table}[!h]
\centering
\caption{Covid Effects (only Sao Paulo state applicants)}\label{tab:effects_SP}
\begin{adjustbox}{max width = \textwidth, width = .\textwidth, center}
\begin{threeparttable}
\begin{tabular}[t]{lccccccc}
\toprule
 && & \multicolumn{5}{c}{Quality Percentiles}\\ \cmidrule(l{10pt}r{10pt}){4-8}
\addlinespace
 Outcome &  Top-100 & Quality& Avg & [0-40] & [40-70] & [70-90] & [90-100] \\ 
 \addlinespace
  & (1) & (2) & (3) & (4) & (5) & (6) & (7) \\ 
\midrule
(Intercept) & \num{0.146} & \num{1.604} & \num{85.825} & \num{0.028} & \num{0.137} & \num{0.217} & \num{0.618}\\
 & (\num{0.004}) & (\num{0.012}) & (\num{0.193}) & (\num{0.002}) & (\num{0.004}) & (\num{0.005}) & (\num{0.006})\\
 \\
Quota & \num{-0.034} & \num{-0.101} & \num{-1.655} & \num{0.009} & \num{0.011} & \num{0.038} & \num{-0.058}\\
 & (\num{0.006}) & (\num{0.019}) & (\num{0.316}) & (\num{0.003}) & (\num{0.006}) & (\num{0.008}) & (\num{0.009})\\
 \\
Covid & \num{0.020} & \num{0.159} & \num{2.246} & \num{-0.009} & \num{-0.034} & \num{-0.015} & \num{0.058}\\
 & (\num{0.006}) & (\num{0.021}) & (\num{0.343}) & (\num{0.003}) & (\num{0.007}) & (\num{0.009}) & (\num{0.010})\\
\\
Num.Obs. & \num{18187} & \num{18187} & \num{18187} & \num{18187} & \num{18187} & \num{18187} & \num{18187}\\
\bottomrule
\end{tabular}
\begin{tablenotes}
    \item \footnotesize{Notes: The table shows the intercept and $\beta$ coefficients of the quota and Covid dummies from the augmented version (\ref{eq:quotas}) for different outcome variables. In the first column, the dependent variable is a dummy variable equal to one if the enrolled student graduated from a top 100 public or top 100 private high schools and zero otherwise. In the second column, the outcome variable is the quality measure defined by equation (\ref{eq:quality}) standardized to have a mean of zero and a standard deviation of one. In the third column, the dependent variable is the quality percentile of each school. In the last four columns, the outcomes are dummy variables equal to one if the high school of origin of the enrolled student belongs to each quality percentile range and zero otherwise. We cluster the standard errors at the individual level. The sample is restricted to students from the Sao Paulo state.} 
\end{tablenotes}
\end{threeparttable}
\end{adjustbox}
\end{table}

\section{Conclusion} \label{section:conclusion}

The Covid-19 pandemic disrupted educational systems around the globe and had large negative effects on achievement levels. Using individual-level data, we show the impact of the Covid pandemic on the transition from high school to a prestigious Brazilian university, Unicamp. 

First, we document a large concentration of Unicamp applicants and enrollees from the top of the high-school quality distribution, using data on the 2016 to 2018 admission processes. We show that most enrollees are drawn from a relatively small set of schools. 

The implementation of a bold affirmative action policy, a quota policy reserving more than one-third of seats for public high-school graduates and underrepresented racial minorities, decreased the concentration at the top by 8-18\% among enrollees admitted in 2019 and 2020. Part of this effect was redirected to students from schools at the bottom of the distribution. As a result, the proportion of students coming from schools in the bottom 40\% increased by 67\%. 

The Covid pandemic reversed most of the deconcentration gains achieved by the quota policy. The concentration at the top raised by 9-13\%. The probability of an enrolled student coming from a high school in the bottom 40\% fell 46\%. We show how these results are entirely driven by students enrolled in high school during the pandemic. 

Thus, our results suggest substantial impacts on the crucial transition from high school to college beyond the documented negative effects of the Covid pandemic on students from different educational levels. More generally, our results suggest a critical role for high-quality schools in helping students persist in the presence of negative shocks, such as the Covid pandemic.

\newpage

\singlespacing
\bibliography{main.bib}

\end{document}